\documentclass{article}
\usepackage{spconf,amsmath,graphicx,graphics,epstopdf}
\usepackage{url}

\newenvironment{itemize*}%
  {\begin{itemize}%
    \setlength{\itemsep}{0pt}%
    \setlength{\parskip}{0pt}}%
  {\end{itemize}}

\title{A COMPARATIVE STUDY OF PITCH EXTRACTION\\ALGORITHMS ON A LARGE VARIETY OF SINGING SOUNDS}
\name{Onur Babacan$^1$\thanks{O. Babacan is supported by a PhD grant funded by UMONS and Acapela Group S.A. GIPSA-lab (UMR5216: CNRS, INPG, Univ. Stendhal, UJF).}, Thomas Drugman$^1$, Nicolas d'Alessandro$^1$, Nathalie Henrich$^2$, Thierry Dutoit$^1$}
\address{$^1$Circuit Theory and Signal Processing Laboratory, University of Mons, Belgium\\
$^2$Speech and Cognition Department, GIPSA-lab, Grenoble, France}
\begin{document}
\ninept
\maketitle
\begin{abstract}

The problem of pitch tracking has been extensively studied in the speech research community. The goal of this paper is to investigate how these techniques should be adapted to singing voice analysis, and to provide a comparative evaluation of the most representative state-of-the-art approaches. This study is carried out on a large database of annotated singing sounds with aligned EGG recordings, comprising a variety of singer categories and singing exercises. The algorithmic performance is assessed according to the ability to detect voicing boundaries and to accurately estimate  pitch contour. First, we evaluate the usefulness of adapting existing methods to singing voice analysis. Then we compare the accuracy of several pitch-extraction algorithms, depending on singer category and laryngeal mechanism. Finally, we analyze their robustness to reverberation.

\end{abstract}
\section{Introduction}
\label{sec:intro}

Over the last decades, research fields associated with speech understanding and processing have seen an outstanding development. This development has brought a diverse set of algorithms and tools for analyzing, modeling and synthesizing the speech signal. Although singing is achieved by the same vocal apparatus, transposing the speech approaches to singing signals may not be straightforward \cite{Kob2011}. In particular, pitch range in singing is wider than in speech, pitch variations are more controlled, dynamic range is greater, and voiced sounds are sustained longer. The impact of source-filter interaction phenomena is also greater in singing than in speech, and thus they can less easily be neglected \cite{Titze2008}. In addition, the diversity in singer categories and singing techniques make it difficult to consider the "singing voice" as a whole and take a systematic analysis approach. As a result, speech and singing research fields have rather evolved side by side, obviously sharing several approaches, but singing research has not encountered the same formalization and standardization as in speech research.

One consequence of such a difficulty to approach the wide range of singing voices as a whole is the lack of singing synthesis techniques that can address such variability. It results in a limited set of singing synthesizers, generally focusing on one singer category or one singing technique. Therefore it remains quite far from expressive abilities of real humans, but also far from concrete needs of musicians wishing to use these tools. Among existing systems, Harmonic plus Noise Modeling (HNM) has been used extensively \cite{Stylianou05}. In SMS \cite{SMS} and Vocaloid\cite{vocaloid}, HNM is used to bring a degree of control over a unit concatenation technique \cite{Bonada01}, though it limits the synthesis results in the range of the prerecorded samples. In CHANT \cite{Rodet84}, FOF \cite{FOF} synthesis has been coupled with a rule-based description of some typical operatic voices, showing remarkable results for soprano voices. Meron has applied the non-uniform unit selection technique to singing synthesis \cite{Meron00}, showing convincing results but only for lower registers. Similar strategies have been applied to formant synthesis, articulatory synthesis \cite{Birkholz06} and HMM-based techniques \cite{Saino06}, with similar limitations in extending the range of vocal expression.

In this research, we make the first step in building an analysis framework, targeting the synthesis of the singing voice for a wide range of singer categories and singing techniques. Indeed we have been working on expressive HMM-based speech synthesis for several years \cite{Astrinaki12,DSM,PicartArticulation} and we now aim to adapt our analysis framework to a wide range of singing voice databases. The purpose of this benchmarking work is to systematically evaluate various analysis algorithms -- which happen to come from speech processing -- among a large reference database of annotated singing sounds and drive some differentiated conclusions, i.e. determine the best choices to make regarding various properties of the singer and the singing technique. Our first study focuses on pitch extraction, as it is among the most prominent parameters in singing analysis/synthesis and it will be used as the foundation for many further analysis techniques. We also decided to discuss the pitch extraction errors among three main properties: singer category, laryngeal mechanism and the effect of reverberation.

The structure of the paper is the following: Section \ref{sec:Methods} briefly describes the pitch trackers that are compared in this study, and investigates what adaptation can be considered to make them suitable for singing voice analysis. Our experimental protocol is presented in Section \ref{sec:Protocol}, along with the database, the ground truth extraction and error metrics. Results are discussed in Section \ref{sec:Results}, investigating the impact of various factors on the performance of pitch trackers. Finally, we narrow down some conclusions to the study in Section \ref{sec:Conclusion}.

\section{Methods for Pitch Extraction}
\label{sec:Methods}

\subsection{Existing Methods}
\label{ssec:ExistingMethods}

In this paper, we compare the performance of six of the most representative state-of-the-art techniques for pitch extraction. They were reported to provide some of the best results to analyze speech signals \cite{SRH}, and are now briefly described.

\begin{itemize}

\item \textbf{PRAAT}: Commonly used in speech research, the PRAAT package \cite{PRAAT} provides two pitch tracking methods. In this paper, we used PRAAT's default technique which is based on an accurate autocorrelation function. This approach was shown in \cite{PRAAT} to outperform the original autocorrelation-based and the cepstrum-based techniques on speech recordings.

\item \textbf{RAPT}: Released in the ESPS package \cite{ESPS}, RAPT \cite{RAPT} is a robust algorithm that uses a multi-rate approach. Here, we use the implementation found in the SPTK 3.5 package \cite{SPTK}.

\item \textbf{SRH}: As explained in \cite{SRH}, the Summation of Residual Harmonics (SRH) method is a pitch tracker exploiting a spectral criterion on the harmonicity of the residual excitation signal. In \cite{SRH}, it was shown to have a performance comparable to the state-of-the-art on speech recordings in clean conditions, but its use is of particular interest in adverse noisy environments. In this paper, we use the implementation found in the GLOAT package \cite{GLOAT}.

\item \textbf{SSH}: This technique is a variant of SRH which works on the speech signal directly, instead of the residual excitation.

\item \textbf{STRAIGHT}: STRAIGHT \cite{STRAIGHT} is a high-quality speech analysis, modification and synthesis system based on a source-filter model. There are two pitch extractors available in the package and we use the more recently integrated one as published in \cite{STRAIGHT2}. This method is based on both time interval and frequency cues, and is designed to minimize perceptual disturbance due to errors in source information extraction.

\item \textbf{YIN}: YIN is one of the most popular pitch estimators. It is based on the autocorrelation method, making several refinements to reduce possible errors \cite{YIN}.  In this paper, we used the implementation freely available at \cite{YINWeb}. 

\end{itemize}

The following section aims at investigating how these techniques can be adapted for the analysis of singing voice.

\subsection{Adapting Pitch Trackers to Singing Voice}
\label{ssec:AdaptingTrackers}

Since the algorithms presented in Section \ref{ssec:ExistingMethods} have been designed and optimized for speech, the set of default input parameters might not be suitable for processing the singing voice. To measure the effect of various parameters, we applied a range of input parameters where available, depending on the algorithm. The main parameter we varied was the window length, as it introduces a trade-off between analyzing low-pitched voices (which requires longer windows encompassing at least two glottal cycles to have a periodicity) and precisely following the pitch contour (which requires shorter windows to capture fine pitch variations). For SRH, SSH and YIN, window length was varied and optimized; with values of 125 ms, 100 ms and 10 ms respectively, in comparison to the respective default values of 100 ms, 100 ms and 16 ms. (SSH happened to use the optimum value by default). As a second parameter, we addressed setting the threshold used for voiced/unvoiced (V/UV) detection. This was applied for PRAAT, SRH and SSH with values of 0.25, 0.065 and 0.095 respectively, in comparison to the default values of 0.45, 0.07 and 0.07. For the purpose of consistency, the F0 search range was set between 60-1500 Hz, to account for the wide vocal range in singing. A 10-ms frame shift was chosen for all methods, with the exception of STRAIGHT. Since the  STRAIGHT algorithm is partially-based on instantaneous frequency, and the default shift interval is 1ms, using 10 ms  caused significant inaccuracies and large jumps in the contour. To compare results to the others, we used the default shift of 1 ms, and downsampled the resultant contour by 10. We also verified the synchronicity of these contours by visually comparing a small but representative set against the corresponding RAPT contours.
 


Covering all combinations of parameters would have required a prohibitively large amount of computation time, consequently we chose to use a two-stage search for the best values. This is an acceptable substitute to complete optimization, since the two considered parameters have different, almost independent effects on the performance.  In this process, we first find the best threshold value at the default window length by minimizing the voicing decision error (see Section~\ref{ssec:ErrorMetrics}). Then, we find the best window length value by minimizing F0 frame error (see Section~\ref{ssec:ErrorMetrics}) at this threshold value.

Additionally, as a complement to the methods described in Section \ref{ssec:ExistingMethods} and their optimized versions, we investigated the usage of a post-processing approach \cite{postfiltering} originally developed for improving YIN results on music data. This post-process makes use of statistical information as well as some musical assumptions to correct sudden changes the F0 contour. Even though not all algorithms are heavily prone to such errors, we applied it to all of them for a fair comparison (see Section~\ref{sec:Results}).

In the cases where reliable voiced/unvoiced decisions were not available, we substituted the decisions from RAPT to calculate error metrics which required them. Specifically, these cases were YIN and STRAIGHT, the former due to YIN not providing these decisions, and the latter due to prohibitively high error rate, making comparisons incompatible, as will be explained further in Section~\ref{sec:Results}

\section{Experimental Protocol}
\label{sec:Protocol}

\subsection{Database}
\label{ssec:Database}
For this study, the scope was constrained to vowels in order to limit the effects of co-articulation on pitch extraction. Samples for 13 trained singers were extracted from the LYRICS database recorded by \cite{Henrich2001,Henrich2005}. The selection comprised 7 bass-baritones (B1 to B7), 3 countertenors (CT1 to CT3), and 3 sopranos (S1 to S3). The recording sessions took place in a soundproof booth. Acoustic and electroglottographic signals were recorded simultaneously on the two channels of a DAT recorder. The acoustic signal was recorded using a condenser microphone (Br\"{u}el \& Kj\ae r 4165) placed 50 cm from the singer's mouth, a preamplifier (Br\"{u}el \& Kj\ae r 2669), and a conditioning amplifier (Br\"{u}el \& Kj\ae r NEXUS 2690). The electroglottographic signal was recorded using a two-channel electroglottograph (EG2, \cite{Rothenberg1992}). 
The selected singing tasks comprised sustained vowels, crescendos-decrescendos and arpeggios, and ascending and descending glissandos. Whenever possible, the singers were asked to sing in both laryngeal mechanisms M1 and M2 \cite{Henrich2006,Roubeau2009}. Laryngeal mechanisms M1 and M2 are two biomechanical configurations of the laryngeal vibrator commonly used in speech and singing by both male and females. Basses, baritones and countertenor singers mainly use M1 for singing, but they also have the possibility to sing in M2 in the medium to high part of their tessitura. Sopranos mainly sing in M2, but they can choose to sing in M1 in the medium to low part of their tessitura.

\subsection{Ground Truth}

In order to objectively assess the performance of pitch trackers, a ground truth (i.e a reference pitch contour) is required. To obtain this, we used the RAPT algorithm on the synchronized electroglottography (EGG) recordings. The choice of RAPT is justified by the fact that it was shown in \cite{SRH} to outperform other approaches on clean speech signals. In addition, we produced pitch contours extracted from both the EGG and the differentiated-EGG (dEGG) signals, and applied a manual verification process by visually comparing each contour to the spectrogram of the EGG signal. We then either selected the better of the two options, or excluded the considered sample from the experiment if both were found to be erroneous in some parts. The resultant experiment database consists of 524 recordings for which we have a reliable and accurate ground truth. 

\subsection{Error Metrics}
\label{ssec:ErrorMetrics}
In order to assess the performance of the pitch extraction algorithms,the following four standard error metrics were used \cite{f0frameerror}: 

\begin{itemize}
\item \textbf{Gross Pitch Error (GPE)} is the proportion of frames, considered voiced by both pitch tracker and ground truth, for which the relative pitch error is higher than a certain threshold (usually set to 20\% in speech studies \cite{SRH}). In this work, we fixed this threshold to one semitone, in order to make the results meaningful from the musical perception point of view. All error calculations are done in the unit of cents (one semitone being 100 cents). 


\item \textbf{Fine Pitch Error (FPE)} is the standard deviation of the distribution of relative error values (in cents) from the frames that do not have gross pitch errors. Both estimated and reference V/UV decisions must then be voiced.

\item \textbf{Voicing Decision Error (VDE)} is the proportion of frames for which an incorrect voiced/unvoiced decision is made.

\item \textbf{F0 Frame Error (FFE)} is the proportion of frames for which an error (either according to the GPE or the VDE criterion) is made. FFE can be seen as a single measure for assessing the overall performance of a pitch tracker.
\end{itemize}

\section{Results}
\label{sec:Results}

Our experiments are divided into four parts. In Section \ref{ssec:AdaptResults}, the need to adapt pitch trackers for the analysis of singing voice is quantified. Sections \ref{ssec:SingerCatResults} and \ref{ssec:LaryngealResults} investigate the effect of singer category (baritone, countertenor, soprano) and laryngeal mechanism on pitch estimation performance. Finally the robustness to reverberation is studied in Section \ref{ssec:Reverb}.

\subsection{Utility of Adapting Pitch Trackers to Singing Voice}
\label{ssec:AdaptResults}

The overall performance of the compared techniques (with their variants) across the whole database is displayed in Table~\ref{tab:allErrors}. To distinguish between the two steps mentioned in Section \ref{ssec:AdaptingTrackers}(parameter optimization and post-processing), an asterisk denotes the post-processed version of the algorithm output, letter v denotes that V/UV decisions from RAPT was used instead of the algorithm's own, and letter u denotes "unoptimized", meaning the results were obtained with default input parameters. Optimization was done on window length and V/UV threshold, for SRH, SSH, YIN, and SRH, SSH, PRAAT, respectively. The effect of optimization is marginal on PRAAT results, however, it is significant on SRH and SSH. This is due in great extent to a proper selection of the window length which results in a noticeable decrease of GPE, as well as slight reduction of FPE. For YIN, we observe a small and acceptable trade-off between GPE and FPE when optimized for GPE.

As mentioned in Section~\ref{ssec:AdaptingTrackers}, V/UV decisions from RAPT are used for all error calculations of STRAIGHT and YIN. Using the V/UV decisions from STRAIGHT, we observed VDE rates higher than 30\% among all data groupings we investigated. While this had the side effect of greatly improving GPE due to selection bias, it was not a consistent comparison to the other methods, thus we completely discarded V/UV decisions from STRAIGHT.

Except for STRAIGHT and PRAAT, it can be observed that applying the post-process yields an appreciable improvement for all other techniques. While maintaining a constant efficiency in terms of voicing decisions, and similar FPE performance, the post-process allows an important reduction of GPE. This is particularly well emphasized for RAPT and YIN algorithms. In the remainder of our experiments, we will always refer to the optimized, post-filtered results from an algorithm as it leads to the best results. 

Comparing the various techniques in Table~\ref{tab:allErrors}, we observe that PRAAT, followed by RAPT, gives the best determination of voicing boundaries. Regarding the accuracy in the pitch contour estimation, RAPT* and YIN* provide the lowest gross error rates, while YIN is clearly seen to lead to the lowest FPE.

\begin{table}[htp]
\caption{Error Rates Across the Whole Dataset}
\begin{center}
\begin{tabular}{|l|c|c|c|c|}
\hline
 & \textbf{GPE (\%)} & \textbf{FPE (C)} & \textbf{VDE (\%)} & \textbf{FFE (\%)} \\ \hline
RAPT &1.01 & 21.96 & 1.05 & 1.99  \\ \hline
RAPT*  &\textbf{0.65} &21.98 &1.05 &\textbf{1.66}  \\ \hline
STRAIGHTv &1.26 &17.22&1.05 &2.22  \\ \hline
STRAIGHTv*  &1.25 &17.22 &1.05 &2.21  \\ \hline
PRAATu  &1.47 & 21.91&\textbf{0.81} &2.18  \\ \hline
PRAAT   & 1.41 &21.93 &\textbf{0.81} &2.15  \\ \hline
PRAAT* & 1.41 &21.94 &\textbf{0.81} & 2.13  \\ \hline
SRHu    &1.91 &18.99 &1.28 &3.08  \\ \hline
SRH    &1.72 &17.33 &1.33 &2.95  \\ \hline
SRH*   &1.61 &17.36 &1.33 &2.84  \\ \hline
SSHu   &3.51 &19.66 &1.27 &4.55  \\ \hline
SSH   &2.40&19.46 &1.39 &3.61  \\ \hline
SSH*   & 1.91&19.43 &1.39 &3.16  \\ \hline
YINvu    & 2.69&\textbf{8.38} &1.05 &3.56  \\ \hline        
YINv    & 2.44&12.79 &1.05 &3.32  \\ \hline        
 YINv*   &0.91 &12.95 &1.05 &1.9  \\ \hline        
\end{tabular}
\end{center}
\label{tab:allErrors}
\end{table}%
\vspace{0mm}




\subsection{Effect of Singer Categories}
\label{ssec:SingerCatResults}

\begin{figure}[htbp] 
   \centering
   \includegraphics[width=\linewidth]{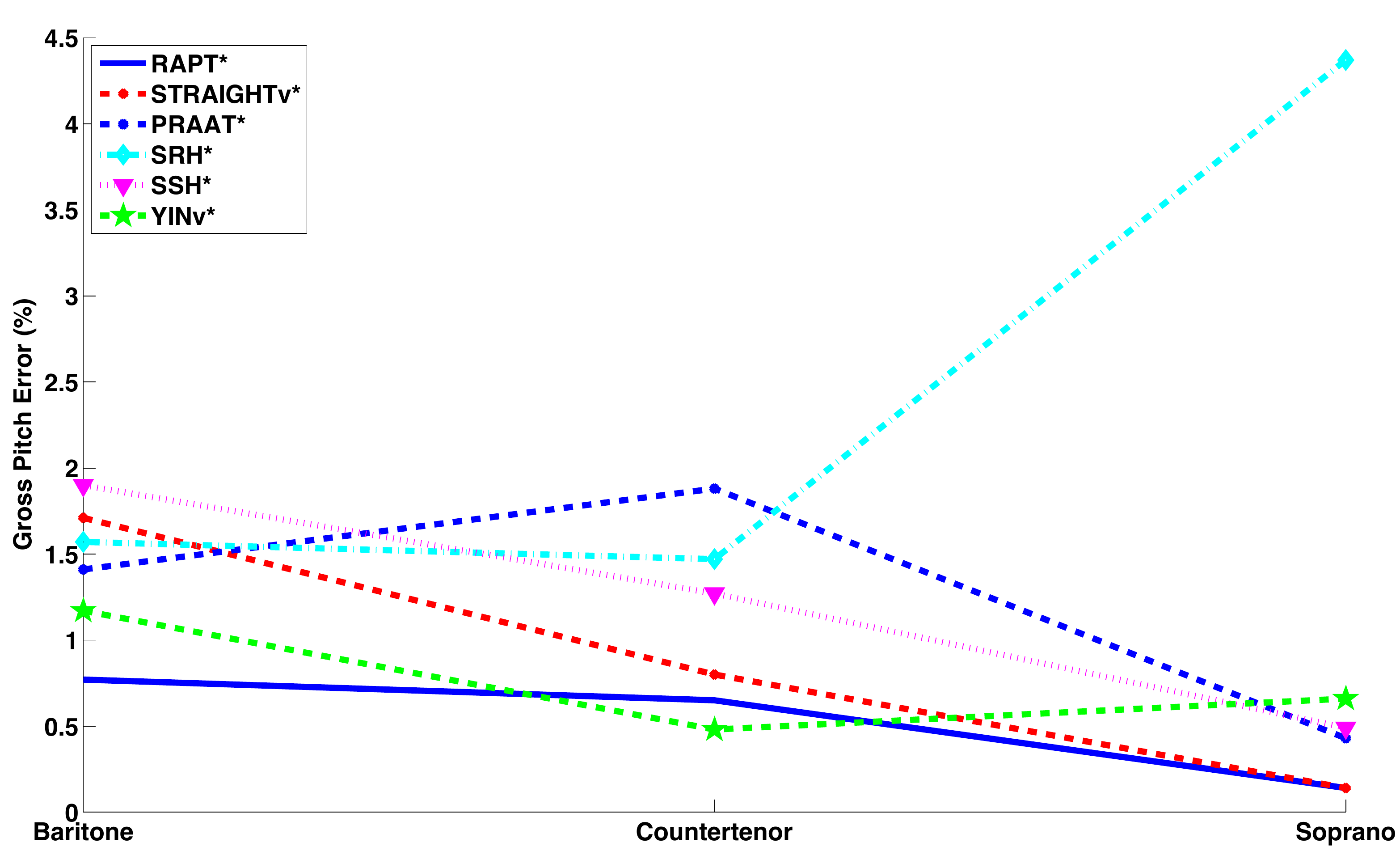} 
   \caption{Effect of Singer Category on Gross Pitch Error (GPE)}
   \label{fig:singerGpe}
\end{figure}

Three categories of singers, characterized by different vocal ranges (indicated hereafter between parentheses as musical notes) are represented in our database: baritones (F2 to F4), countertenors (F3 to F5), and sopranos (C4 to C6). The effect of the singer category on the GPE, which should reflect the pitch range differences, is given in Figure \ref{fig:singerGpe}. Except for SRH which suffers for a dramatic degradation for sopranos, the performance of all other techniques follow the same trends: GPE decreases as the vocal range goes towards higher pitches. Going from baritones to sopranos, GPE is observed to be divided by a factor between 2 and 4, depending on the considered technique. Our results on FPE revealed similar conclusions: for all methods, the standard deviation of the relative pitch error distribution decreases from baritones to sopranos. This reduction varies between 2 and 7 cents across algorithms, with the best performance achieved by YIN* (15 cents for baritones, and 8.4 cents for sopranos).

\vspace{0mm}


\subsection{Effect of Laryngeal Mechanisms}
\label{ssec:LaryngealResults}

\begin{figure}[htbp] 
   \centering
   \includegraphics[width=0.85\linewidth]{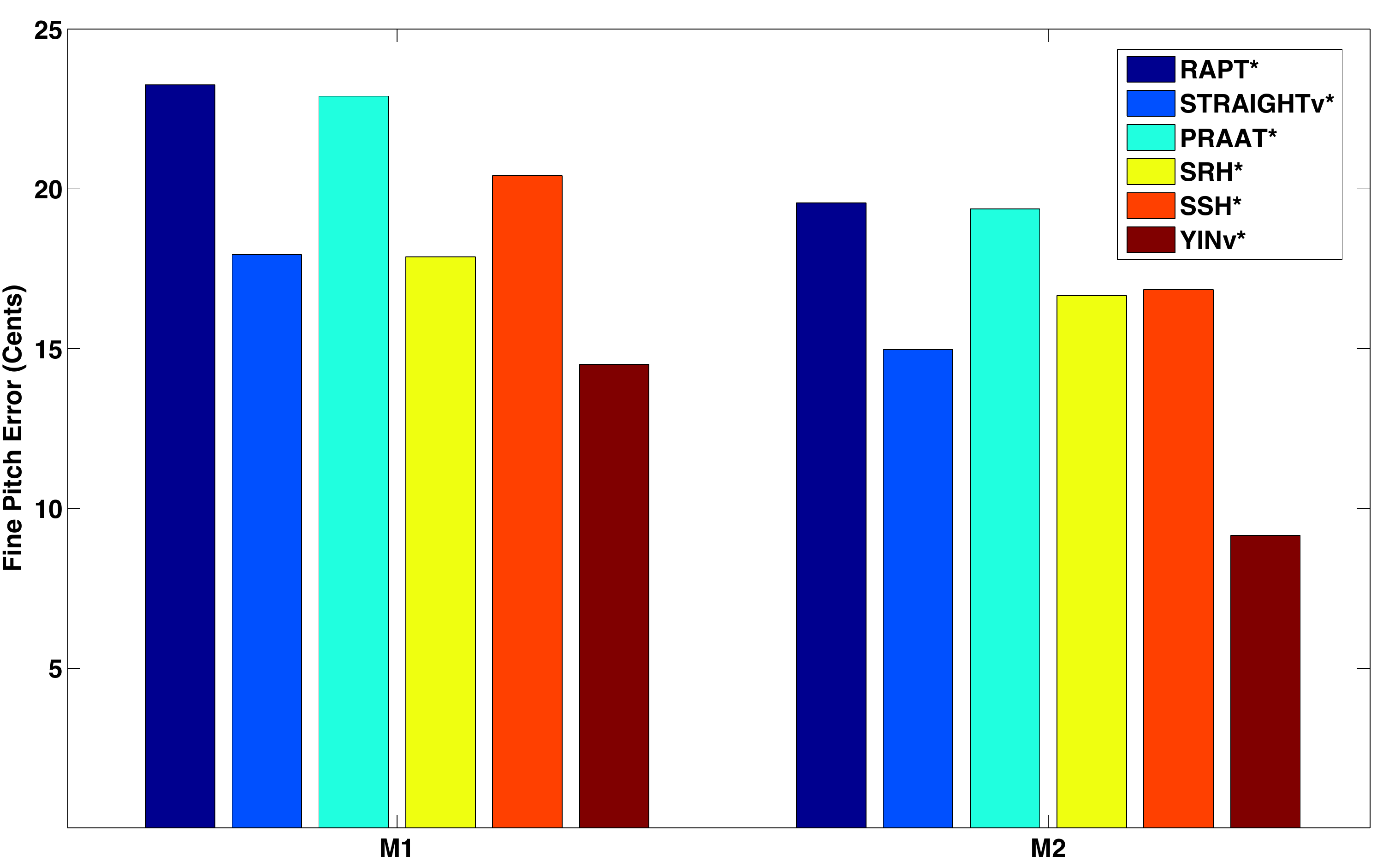} 
   \caption{Effect of Laryngeal Mechanism on Fine Pitch Error (FPE)}
   \label{fig:mechanismFpe}
\end{figure}

Laryngeal mechanisms used by singers have been described in Section \ref{ssec:Database}. We now inspect what the influence of these mechanisms is on the efficiency of the compared pitch estimation techniques. The impact on FPE is illustrated in Figure \ref{fig:mechanismFpe}. Again, it is observed that YIN* provides the best FPE results. Consistently across all algorithms, M2 is noticed to lead to lower FPE values. This actually corroborates our findings on the singer category: FPE performance improves as the pitch increases. In the same way, the conclusions we have drawn in Section \ref{ssec:SingerCatResults} for GPE are also observed here\footnote{\label{fn:nospace}Figure omitted due to space constraints.}: M2 is characterized by lower GPE values for all methods except SRH* (whose results for M2 are the worst by a significant margin).


\subsection{Robustness to Reverberation}
\label{ssec:Reverb}

In many concrete cases, singers are placed within large rooms or halls, where the microphone might capture replicas of the voice sound stemming from reflections on the surrounding walls or objects. To simulate such reverberant conditions, we considered the $L$-tap Room Impulse Response (RIR) of the acoustic channel between the source to the microphone. RIRs are characterized by the value $T_{60}$, defined as the time for the amplitude of the RIR to decay to -60dB of its initial value. A room measuring 3x4x5~m and $T_{60}$ ranging \{100, 200,~\ldots,~500\}~ms was simulated using the source-image method~\cite{Allen1979} and the simulated impulse responses convolved with the clean audio signals.

\begin{figure}[htbp] 
   \centering
   \includegraphics[width=0.95\linewidth]{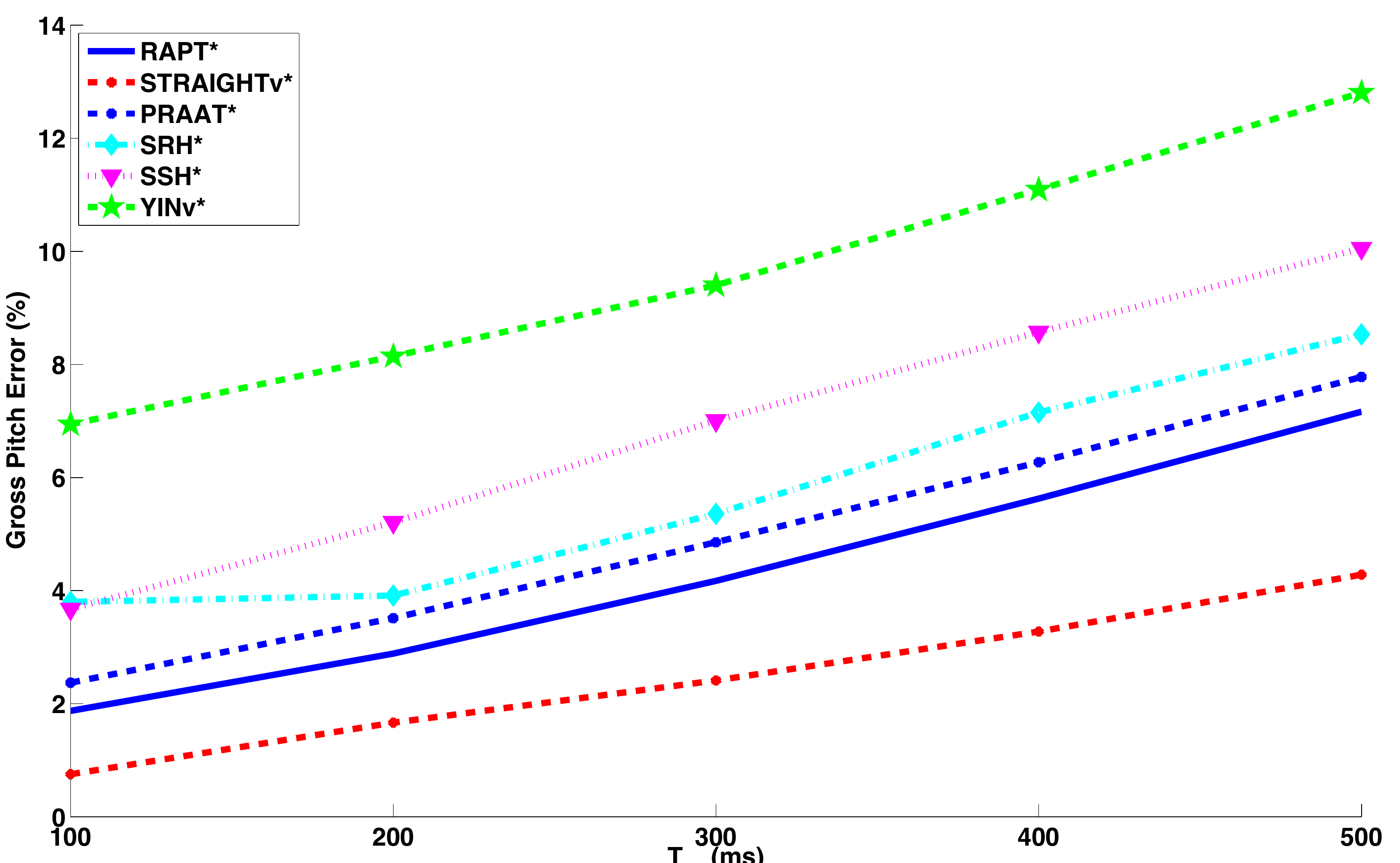} 
   \caption{Effect of Reverberation on Gross Pitch Error (GPE)}
   \label{fig:reverbGpe}
\end{figure}

Results of GPE as a function of the level of reverberation are presented in Figure \ref{fig:reverbGpe}. Even in the less severe condition (i.e. when $T_{60}$ is 100 ms), the performance of pitch estimation techniques is observed to be affected (these results are to be compared with those reported in Table \ref{tab:allErrors} for non-reverberant recordings). More particularly, YIN* suffers from the most important degradation: with a GPE of 0.91\%, it now reaches a value around 7\%. In contrast, STRAIGHTv* turns out to be the most robust as it keeps almost the same GPE as in the clean conditions. Regarding their evolution with the reverberation level, all techniques exhibit a similar behavior, with an increase of GPE between 3 and 6\% as $T_{60}$ varies from 100 to 500 ms.

The impact of reverberation on FPE is also examined. Although all techniques but STRAIGHTv* were found to suffer from a substantial increase of GPE even when $T_{60}$ is 100 ms, the effect on FPE is much less pronounced. At that level, we observed that pitch estimators have their FPE increasing by 3 to 5 cents, which is relatively minor; with the exception of YIN*: shown to exhibit the strongest degradation in terms of gross pitch errors, here, it reaches the best accuracy. Regarding their evolution with the reverberation degree, all methods behave very similarly with an increase of FPE between 9 and 13 cents as $T_{60}$ goes from the slightest to the strongest degradation. As a conclusion; even though some techniques (especially YIN*) produce a much higher number of gross errors in reverberant environments, it seems that their ability to precisely follow the pitch contour (when no gross error is made) is rather well preserved.

\section{Conclusion}
\label{sec:Conclusion}
As a first step towards developing efficient techniques of singing voice analysis and synthesis, this paper provided a comparative evaluation of pitch tracking techniques. This problem has been addressed extensively for the speech signal, and the goal of this paper was to answer two open questions: \emph{i}) what adaptation is required when analyzing singing voice?, and \emph{ii}) what is the best method to extract pitch information from singing recordings? Six of the most representative state-of-the-art methods were compared on a large dataset containing a rich variety of singing exercises. As an answer to question \emph{i}, both the use of parameter settings specific to singing voice and post-processing of pitch estimates led to an appreciable reduction of gross pitch errors. The answer to question \emph{ii} depended on the considered error metric. PRAAT and RAPT provided the best determination of voicing boundaries. RAPT reached the lowest number of gross pitch errors. YIN achieved the best accuracy. Pitch-estimation performances were better for sopranos than for baritones and counter tenors, and for singers in laryngeal mechanism M2. Finally, the robustness of the techniques in reverberant conditions was studied, showing that YIN suffered from the strongest degradation, while STRAIGHT was the most robust.

\newpage
\bibliographystyle{IEEEbib}
\bibliography{strings,refs}

\end{document}